\documentclass[sn-mathphys]{sn-jnl}

\jyear{2021}%

\theoremstyle{thmstyleone}%
\theoremstyle{thmstyletwo}%

\theoremstyle{thmstylethree}%

\usepackage{graphicx}
\usepackage{bm}
\usepackage{xcolor}

\begin{document}

\newcommand{\be}{\begin{equation}}
\newcommand{\ee}[1]{\label{#1}\end{equation}}
\newcommand{\bem}{\begin{eqnarray}}
\newcommand{\eem}[1]{\label{#1}\end{eqnarray}}
\newcommand{\eq}[1]{Eq.~(\ref{#1})}
\newcommand{\Eq}[1]{Equation~(\ref{#1})}

\newcommand{\rc}[1]{\textcolor{red}{#1}}

\title[Joe Vinen and transverse force on vortex]{Joe Vinen and transverse force on vortex}


\author*{\fnm{Edouard} \sur{Sonin}}\email{sonin@cc.huji.ac.il}

\affil{\orgdiv{Racah Institute  of Physics}, \orgname{Hebrew University of Jerusalem}, \orgaddress{\street{Givat Ram}, \city{Jerusalem}, \postcode{9190401}, 
 \country{Israel}}}


\abstract{The paper gives a 
 glimpse on the problem of transverse force on a vortex,  its proper determination in general and the role in the process of quantum vortex nucleation. Investigation of this problem is an essential part of the scientific heritage of Joe Vinen as an experimentalist and a theoretician.}




\maketitle

\section{Introduction}\label{sec1}

The Magnus force on a vortex has long been known in classical hydrodynamics \cite{Lam}. This force appears if the vortex
moves with respect to a liquid. The force is normal to the relative vortex velocity and therefore is reactive and does not
produce a work. In general, such a force arises always when a body with a flow circulation around it moves through a
liquid or a gas (the Kutta-Joukowski theorem). The most important example is the lift force on a wing of an airplane
which keeps the airplane in the air \cite{LL}.

The key role of the Magnus force  was clear from the seminal paper of Hall and Vinen \cite{Hal56b} , which marked  the beginning of investigations of superfluid  vortex dynamics and the emergence of the theory, which now is called the Hall--Vinen--Bekarevich--Khalatnikov (HVBK) theory.
The Magnus force is proportional to the quantum of the velocity circulation around  the vortex. This was used by Joe Vinen  in his classical experiment \cite{VinQ} on the first detection of the circulation quantum in observations of vibrations of a fine wire with a trapped vortex line in superfluid $^4$He. The existence of the Magnus force in type II superconductors was demonstrated theoretically  by Nozie\`res and Vinen \cite{NozVin}.

The superfluid Magnus force was defined as a force between a vortex and a superfluid and was proportional to the superfluid density. But in the two-fluid hydrodynamics the Magnus force is not the only transverse force on the vortex: there
was also a transverse force produced by quasiparticles
moving past the vortex. The transverse force
from rotons was found by Lifshitz and Pitaevskii \cite{Lif57} from the
semiclassical scattering theory. Later Iordanskii  \cite{Ior5} revealed
the transverse force from phonons which was equal in magnitude
and opposite in sign with the force calculated by 
Lifshitz and Pitaevskii. 

From the very beginning the Iordanskii
force was a subject of controversy. Iordanskii  suggested
that his force and the Lifshitz-Pitaevskii force were of different
origins and for rotons they should be summed. As a
result, he concluded that the transverse force from rotons
vanished. Later it was  demonstrated \cite{Son75} that
the Iordanskii force for rotons is identical to the Lifshitz-
Pitaevskii force and they must not be added. In addition, the
Lifshitz-Pitaevskii force from rotons was calculated in the
original paper \cite{Lif57} with a wrong sign. After its correction the
transverse force on the vortex had the same sign and magnitude both for rotons (the Lifshitz-Pitaevskii force) and for
phonons (the Iordanskii force) \cite{Son75,PRB7,Son10}.  The transverse force from quasiparticles
results from interference between quasiparticles which move
past the vortex on the left and on the right sides with different
phase shifts, like in the Aharanov-Bohm effect \cite{AB}.  In clean
superconductors the BCS quasiparticles produce an additional
transverse force on the vortex \cite{Gal,Kop76b} analogous to the
transverse force from quasiparticles in superfluids. 

The controversy around the transverse force on the vortex was revived after the paper of Ao and Thouless \cite{AT}. They came to the conclusion that there is a universal exact expression for the total transverse force on the vortex derived from the concept of the geometrical phase (the Berry phase). The total transverse force  coincided with the superfluid Magnus force and was proportional to the superfluid density $\rho_s$. According to Ao and Thouless, there was no transverse force on the vortex from quasiparticles and impurities.

The Ao-Thouless theory was in evident disagreement
with the previous calculations of the transverse force on the
vortex  in superfluids and superconductors. It attracted a great attention and launched a vivid discussion \cite{HHWex,Wex,ComAT,Wex2}. Joe Vinen had a great interest to this dispute. In  fact, he was a moderator and discussed the issue with those involved in the dispute.
Eventually  with his  participation some consensus  was
reached that the original calculation of the Berry phase missed the
contribution from the normal-fluid circulation  \cite{Th,Magn}.
This rehabilitated transverse forces from quasiparticles and impurities.

Considering the problem of critical velocities in superfluids Vinen \cite{Vin61,Vinen63} attracted  attention to the fundamental problem of quantum nucleation of vortices. It was known from classical hydrodynamics that the energy of a vortex ring in a moving ideal fluid is not monotonous. At small ring radius  the energy grows but at large ring radius it decreases and becomes negative. According to the Landau criterion for superfluids, this should mean that the superfluid flow is unstable. Vinen stressed an essential difference between elementary excitations (phonons, rotons), for which the Landau criterion was suggested, and the vortex ring. The latter is a macroscopic  excitation. Its generation is accompanied by changing the states of a macroscopic number of particles. As a result, although the vortex nucleation is not forbidden by the energy conservation law, its probability is expected to be very low and requires an estimation. At zero temperature the vortex nucleation is a process of quantum tunneling through  the potential barrier separating vortex rings of small and large radius. The transverse force on a vortex is important for this process. The physical picture of quantum nucleation crucially depends on whether dynamics of vortex is governed by the inertia force like in Newton's  law, or by the transverse force.

\section{Equation of motion for the vortex} \label{MFMH}

In the hydrodynamics of ideal classical fluids the equation of motion of  the vortex with the circular velocity field around it, 
\be
\bm v_0={\kappa[ {\bm z}\times \bm r]\over 2\pi r^2},
   \ee{}
  is 
\be
-\rho \kappa[ {\bm z}  \times (\bm  v_L-\bm v)]={\bm f}.                         
         \ee{MagCl}
Here $\bm v_L$ is the velocity of the vortex, $\rho$ is the fluid mass density, $\bm v$ is the velocity of the fluid flow past the vortex, $\bm f$ is an external force per unit length of the vortex,  $\kappa$ is the circulation of the velocity $\bm v_0$, and $\bm z$ is the unit vector along the  $z$ axis (axis of the vortex). All  other vectors are in the plane normal to the $z$ axis. In the absence of external forces the vortex moves with the transport fluid velocity $\bm v$  (Helmholtz's theorem).

In the two-fluid hydrodynamics of superfluids  $\kappa =h/m$ is the circulation quantum, and
the stationary vortex motion is described by the equation
\be
-\rho_s\kappa[ {\bm z} \times (\bm  v_L-\bm v_s)] -{\bm f}_{fr} ={\bm f},
                     \ee{eq2fl}
where 
\be
{\bm f}_{fr}= -D'[{\bm z} \times (\bm  v_L-\bm  v_{nl})]-D(\bm  v_L-\bm  v_{nl}) 
                     \ee{fr}
is the mutual friction force,   $\rho_s$ and $\rho_n$   are the mass densities of the superfluid and normal components respectively, $\bm v_s$ is the superfluid velocity, and $\bm v_{nl}$ is the local normal velocity, i.e., the velocity close to the vortex. The forces proportional to $D$ and $D'$ arise from scattering of quasiparticles by the vortex. 

The  force $\propto D'$ is the transverse Iordanskii force. The theory of scattering of non-interacting phonons and rotons  yields that $D'=-\rho_n \kappa$ \cite{Son75,EBS}. Then in the absence of the external force $\bm f$ and neglecting the dissipative force $\propto  D$ the vortex moves not with the superfluid velocity but with the center-of-mass velocity ${\rho_s\over \rho}\bm v_s+{\rho_n\over \rho}\bm v_n$. This is a generalization of Helmholtz's theorem for superfluids. But the value $D'=-\rho_n \kappa$ is not universal and can be invalid at temperatures close to critical or  in dirty Fermi superfluids  \cite{EBS,Kop}. Note that while in classical hydrodynamics and sometimes in superfluid hydrodynamics the term ``Magnus force'' refers to the whole first term in \eq{eq2fl}  proportional   to the relative velocity 
$\bm  v_L-\bm v_s$, in the theory of superconductivity only the part of this term proportional to the vortex velocity  $\bm  v_L$ is called Magnus force. The other part proportional to the superfluid velocity  $\bm  v_s$ is called the Lorentz force  \cite{Kop}. In this paper we use the latter definition of the Magnus and Lorentz force.

The difference between the local normal velocity  $\bm v_{nl}$ close to the vortex and the normal velocity $\bm v_n$ very far from the vortex,
\begin{equation}
 \bm  v_n  - \bm v_{nl}= \frac{\ln(r_m/r_l)}{4\pi \rho_n\nu}\bm  f_{fr},
                       \label{10.3}
\end{equation}
 was known in the hydrodynamics of classical viscous fluids \cite{Lam}. Here  $\nu$ is the kinematic viscosity. Hall and Vinen \cite{Hal56b} took it into account in their pioneer paper and called this effect viscous drag.
The lower cut-off $r_l$ in the logarithm is usually chosen to be of the order of the mean free path  of quasiparticles.
The upper  cut-off $r_m$ for a single vortex will be defined below. 

Excluding $\bm v_{nl}$ from Eqs.~(\ref{fr}) and (\ref{10.3}) one obtains that the mutual friction force is
\be
{\bm f}_{fr}= -{\cal D}'[{\bm z} \times (\bm  v_L-\bm  v_n)]-{\cal D}(\bm  v_L-\bm  v_n),
                     \ee{frn}
where coefficients ${\cal D}$ and ${\cal D}'$ are connected with  coefficients $D$ and $D'$ by the  complex relation
\begin{equation}
{1\over {\cal D} +i {\cal D}'}={\ln(r_m/r_l)\over 4\pi \rho_n \nu}
+ {1\over D+i D'}.
   \label{DD}  \end{equation}

One can rewrite the equation of vortex  motion \eq{eq2fl}  as 
\be
-\rho_M\kappa[ {\bm z} \times \bm  v_L]+ {\cal D}\bm  v_L=\tilde{\bm f},
                     \ee{rhoM}
where $\rho_M= \rho_s-{\cal D}'/\kappa$ is the effective density, which determines the total  transverse force on the vortex, and the effective force 
\be
\tilde{\bm f}=\bm f-\rho_s\kappa[ {\bm z} \times \bm v_s] +{\cal D}'[{\bm z} \times \bm  v_n)]+{\cal D}\bm  v_n
    \ee{genF}
includes not only the external force $\bm  f$ but also forces produced by motion of the superfluid (the Lorentz force) and the normal components past the vortex.

Equations (\ref{rhoM}) and (\ref{genF}) are valid for Galilean invariant superfluids when the force balance can be determined only by the relative velocities $\bm  v_L-\bm  v_s$ and $\bm  v_L-\bm  v_n$. At zero temperature there is no quasiparticles, and  $\rho_M$ does not differ from $\rho_s= \rho$, since ${\cal D}'=0$.

 In dirty superconductors interaction of Andreev bound states in the vortex core with impurities breaks Galilean invariance and produces the transverse force, which depends only on $\bm v_L$ (Kopnin-Kravtsov force  \cite{Kop76b,Kop,EBS}). The Kopnin-Kravtsov force has a direction opposite  to the superfluid Magnus force, and can  essentially decrease the total transverse force $\propto \rho_M$. The transverse force can be suppressed not only by disorder in dirty superconductors but also in superfluids put into a periodic potential. For example, in the Josephson-junction array the transverse force on a vortex vanishes, and $\rho_M=0$. This  follows from the  particle-hole symmetry \cite{PRB7}. The  transverse force is also suppressed in BEC of cold atoms in  an optical lattice \cite{Son16}.

In accordance with Newton's third law,  the force on the vortex is accompanied by a force $-\bm  f_{fr}$ of opposite direction on the normal fluid. The latter force produces the momentum flux in the normal component, which transports the momentum transferred to the normal component by the force to large distances from the vortex. The total momentum flux  through a cylindric surface $S$ around the vortex equal to $-\bm  f_{fr}$ is:
\bem
-f_{fr\,i} =\oint \left[-{\rho_n\lvert\bm  v_n-\bm  v_L\rvert^2\over 2}\delta_{ij}+\rho_n v(v_{ni}-v_{Li})(v_{nj}-v_{Lj}) - \tau_{ij} \right]dS_{j}.
     \eem{MomFl}
Here 
\be 
\tau_{ij}= \rho_n \nu(\nabla_i v_{nj}+\nabla_j v_{ni})
   \ee{}
is the viscosity tensor, and $dS_{j}$ are components of the vector normal to the surface $S$. Its   modulus is the differential  $dS$ of the surface area . The momentum flux in \eq{MomFl} includes only contributions  connected with the motion of the normal component in the coordinate frame moving with the vortex velocity $\bm v_L$.

At distances from the vortex smaller than the Oseen length  \cite{Lam,EBS}
\be
r_O\sim \nu /\lvert\bm  v_n- \bm  v_L\rvert,  
     \ee{osee}
nonlinear terms quadratic in $\bm v_n-\bm v_L$  can be neglected compared with the viscosity tensor. The Oseen length determines the upper cut-off  $r_m$ in \eq{DD} for a single vortex moving  with constant velocity. 

At distances larger than the Oseen length the nonlinear term becomes a dominant term everywhere excepting the narrow area inside the laminar wake formed behind the vortex moving through the viscous normal fluid. The laminar wake and the viscosity-governed momentum flux inside it are important only for the dissipative longitudinal force proportional to $\cal D$ \cite{EBS}. Outside the laminar flow the momentum flux does not differ from that in a perfect fluid without viscosity, and the transverse force  on the  normal liquid inevitably requires the existence of the circulation of the normal velocity:
\begin{equation}
\kappa_n= \oint d\bm  l \cdot \bm  v_n=- {{\cal D}' \over \rho_n}.
    \label{circ}
      \end{equation}

Ao and  Thouless \cite{AT} connected the total transverse force with the Berry phase proportional the circulation $\oint (d\bm l \cdot \bm  j )$ of the total current $\bm j=\rho_s\bm v_s+ \rho_n\bm v_n $ past the vortex at large distance from the vortex \cite{GWT}.This means that the parameter $\rho_M$ in \eq{rhoM} must be
proportional to the total current circulation:
\be 
\rho_M\propto \oint (d\bm l \cdot \bm  j ).
     \ee{jL} 
In the original version of the theory it was assumed that the velocity circulation is possible only in the superfluid component, while $\kappa_n=0$. This assumption led to the conclusion that the transverse force on the vortex reduces to only the superfluid Magnus force $\propto \rho_s$ and all other forces including the Iordanskii force and the Kopnin-Kravtsov force are ruled out.
Later Thouless {\it et al.} \cite{Th} accepted the existence of the circulation of the normal velocity at large distances from the vortex. This revision has eliminated the disagreement between the Berry phase approach and the previous theories based on the momentum (force) balance investigation. 

According to \eq{DD}, at very large Oseen length ($r_m \sim r_O \to \infty$) the force on the vortex ceases to depend on forces produced by quasiparticle scattering  at the vortex. This refers both to the dissipative longitudinal force ($\propto D$) and the reactive transverse Iordanskii force  ($\propto D'$). Then the only transverse force is the Magnus force, the vortex velocity $\bm v_L$ does not differ  from the local normal velocity $\bm v_{nl}$, and according to \eq{frn}, the friction force is nothing else but the Stokes force on a cylinder moving through a viscous fluid \cite{Lam}:  
\be
{\bm f}_{fr}= -{4\pi \rho_n \nu\over \ln(r_m/r_l)}(\bm  v_L-\bm  v_n),
                     \ee{St}
This regime of vortex motion was assumed by Matheiu and Simon \cite{MS} to interpret experimental data on mutual friction parameters at intermediate temperatures. Since the Oseen length depends 
 on the relative  velocity $\bm  v_L-\bm  v_n$ and diverges at $\bm  v_L-\bm  v_n\to 0$, there is no friction force in this limit. This is  the Stokes paradox \cite{Lam,EBS}. But the upper limit $r_m$ in the logarithm determining the viscous drag is to be the Oseen length only for strictly stationary motion of  a single vortex. In  the oscillatory motion with the finite frequency $\omega$ $r_m$ is   the viscous penetration depth $\sqrt{\nu/\omega}$, and in the case of vortex lattice $r_m$ is the intervortex distance, if these scales are less than the Oseen length (see further details in the book \cite{EBS}).

\section{Equation of motion for the vortex with  mass} \label{VorMass}

Up to now we neglected the inertia force proportional to the vortex acceleration. Taking into account  this force the equation of vortex motion \eq{rhoM} becomes
\be
\mu_v{d\bm v_L\over dt}-\rho_M\kappa[ {\bm z} \times \bm  v_L] =\tilde{\bm f},
                     \ee{rhoMm}
where $\mu_v$ is the  mass per unit length of the vortex.  This equation is analogous to the equation of motion of a charged particle in a magnetic field \cite{Muri}. Here we neglect dissipative forces.

Within the framework of hydrodynamics the effect of the vortex mass is normally very weak, especially in superfluid $^4$He, where the core radius $r_c$ does not exceed a few angstroms.\footnote{See discussion of vortex masses of various origins in the book \cite{EBS}.}  But an important exception from this rule is a vortex line trapped by a flexible wire. This case was realized in Vinen's  famous experiment \cite{VinQ}, which provided the first experimental confirmation of quantization of circulation of a single vortex line. In the experiment oscillations of a thin wire with a trapped vortex line were observed. The wire was coaxial with a cylindric container filled by superfluid $^4$He. The wire together with the trapped vortex line can be considered as a complex vortex line. The radius of the core is  now the radius of the wire $r_w$, which even for a thin wire  is by many orders  larger than the microscopically small core radius $r_c$ of a free vortex.  The mass per  unit length $\mu_v=\pi r_w^2 (\rho_w  +\rho)$ of the solid core includes the mass of the wire itself with mass density $\rho_w$ and the associated mass of the fluid dragged by the wire (the term $\propto \rho$). 

In Vinen's experiment the external force was   a line tension force  $\tilde{\bm f} = -{\cal K} \bm u$ restoring the original axial location of the wire, where ${\cal K}$ is the elastic constant, $ \bm u$ is the two-dimensional vector of displacement in the middle of the wire, and $\bm v_L=d\bm u/dt$. If there is no suppression of the Magnus force at zero temperature, i.e., $\rho_M=\rho$, 
one can rewrite the equation of motion \eq{rhoMm}  as two equations for Cartesian variables $u_x$ and $u_y$ for a monochromatic oscillation mode $\bm u \propto e^{-i\omega t}$:
\bem
-\omega ^2 \mu_v  u_x -i\omega\rho \kappa   u_y=-{\cal K} u_x,
\nonumber \\
-\omega ^2 \mu_v  u_y+i\omega \rho \kappa   u_x=-{\cal K}  u_y.
   \eem{wireOsc}
The dispersion relation for oscillation of the wire is
\begin{equation}
(\omega^2-\omega_0^2)^2 -{\rho^2 \kappa^2 \over \mu_v^2} \omega^2=0,
      \label{DR-w}
\end{equation}
where $\omega_0=\sqrt{{\cal K}/\mu_v}$ is the oscillation frequency of the wire without the Magnus force, when the wire has two  degenerate  modes, which can be chosen as either two
linearly or two circularly polarized modes. In Vinen's experiment there was a weak  Magnus force, which  lifted the degeneracy, and  the dispersion relation yielded  two circularly polarized modes  with two close but different frequencies:
\begin{equation}
\omega=\omega_0\pm {\rho\kappa \over 2 \mu_v}.
      \end{equation}
The presence of two close modes  means that oscillations of the wire are accompanied by beats with
frequencies $\Delta \omega = {\rho \kappa/  \mu_v}$. Measurements of the beat frequency yielded the value of the velocity circulation quantum $\kappa$. Vinen's experiment on circulation measurement in superfluid $^4$He was later repeated in superfluid $^3$He-{\sl B} \cite{PackQ}.

\Eq{wireOsc}  describing oscillations of the wire with
quantum circulation around it illustrates the crossover from  dynamics   governed by the inertia force to dynamics governed by the
Magnus force. If the vortex mass $\mu_v$ decreases and the ratio $\rho \kappa / \mu_v$ becomes very large, the dispersion relation
(\ref{DR-w}) yields two frequencies:
\begin{equation}
\omega_1 ={{\cal K}\over \rho \kappa},\qquad \omega_2 = {\rho \kappa \over \mu_v}.
       \end{equation}
At $\mu_v \to 0$ the frequency of the second mode grows to infinity and cannot be treated within
the hydrodynamical approach.  Only a single circularly polarized mode remains, which is governed by the Magnus force. This  illustrates  the crossover from dynamics of a particle governed by Newton's second  law without transverse forces to  dynamics of  a massless vortex.  In the equation of motion \eq{rhoM} of a massless vortex the force determines not an acceleration but a velocity. Thus, the velocity cannot be an independent variable determined by initial conditions as in the case of a particle. Therefore, a particle in the two-dimensional space has twice a  number of degrees of freedom of a massless vortex performing two-dimensional motion.

\section{Quantum nucleation of a massless vortex}

Vortex nucleation is possible due to thermal or quantum fluctuations. Approaching zero temperature, thermal nucleation of vortices is more and more improbable,  and quantum vortex nucleation becomes predominant.  The vortex is a macroscopic
perturbation of a fluid, and its quantum nucleation is a process of
{\em macroscopic quantum tunneling},  which changes   states of a huge number of particles. The
central assumption of the macroscopic quantum tunneling concept is
that this many-body process can be reduced  to dynamics of one or a few macroscopic degree of freedom. Here we
restrict ourselves to an elementary theory of macroscopic tunneling,
putting aside such an interesting topic as the effect of dissipation
\cite{CL}.

The semiclassical quantum tunneling theory considers motion of a particle in a classically forbidden area under the barrier by transition to imaginary time or coordinate \cite{LLqu}.  One needs to calculate the action  ${\cal S}=\int {\cal L}\,dt$ along the trajectory crossing the classically inaccessible underbarrier region. Here ${\cal L}$ is the Lagrangian. The exponent $\Gamma$ of the  probability of tunneling $W \sim e^{-\Gamma}$ is determined by the imaginary part of the action: $\Gamma = 2\mbox{Im}{\cal S}/\hbar$.  The action along the trajectory follows from the Hamilton--Jacobi theory:
\begin{equation}
{\cal S} =\sum_i \int\limits_L P_i dx_i,
    \end{equation}
where summation is over all pairs of conjugate variables $(x_i,P_i)$ and integration is over the trajectory $L$ determined from the equations of motion.

For application of this procedure to a massless vortex, it is important that the dynamics of this vortex is essentially different from that of a particle governed by Newton's second law: any external force on the vortex is opposed not by the inertia force, but by the superfluid Magnus force. The Magnus force is responsible for the Hall effect  in superconductors, and this type of  tunneling is sometimes called ``Hall tunneling'' \cite{BlatRev}. Since we deal with neutral superfluids we shall call tunneling of a massless vortex Magnus tunneling.

The first analysis of Magnus tunneling for a superfluid vortex was done  by Volovik  \cite{Vol72}. He considered  nucleation of a circular vortex half-loop near a plane boundary.  
Here we consider a straight vortex in a thin film with the superfluid moving along the film edge parallel to the axis $x$ \cite{Son95,EBS}. The vector equation of motion \eq{rhoM} may be considered as the Hamiltonian equations for the pair of conjugate variables ``coordinate $x$  - momentum $P_x$''
\be
{dx \over dt} ={\partial H\over \partial P_x},~~{dP_x \over dt} =-{\partial H\over \partial  x},
   \ee{}
where the momentum of the vortex at the distance $y$ from the edge is
\be
P_x=\rho \kappa y. 
     \ee{}
The Hamiltonian for the superfluid moving with the velocity $v$ is
\be
H=E_v+V(x,y),~~E_v(y)={\rho \kappa^2 \over 4\pi}\ln {y\over r_c} - vP_x,
    \ee{Ham}
where $V(x,y)$  is the potential energy produced by a defect on the film edge and  $E_v$ is the energy of the vortex at the distance $y$ from the film edge parallel to the axis $x$. In contrast to previous sections, in this and next sections density is a mass per unit area of the film, and forces and energies are values for the whole vortex, but not its unit length.

\begin{figure}[h]%
\centering
\includegraphics[width=0.9\textwidth]{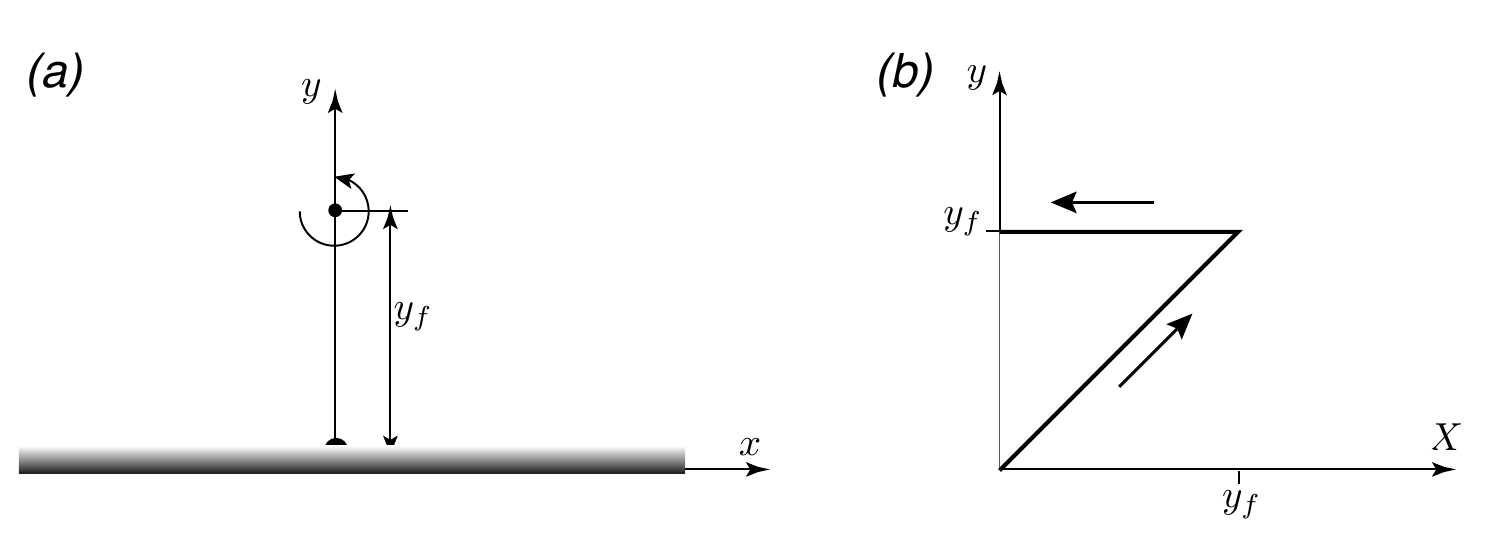}
\caption{Nucleation of a vortex near a film edge with a defect. (a) The vortex nucleated at the defect located at $x=y=0$. After quantum tunneling the vortex appears at $x=0$, $y=y_f$. (b) The trajectory of quantum tunneling in the plane with axes $X=ix$ and $y$. The trajectory starts at $x=X=y=0$, goes along the line $y=X$ until $X=y_f$, and continues along the line $y=y_f$ reaching the final vortex position at $x=0$, $y=y_f$. }\label{fig1}
\end{figure}

In this section we  consider Galilean invariant superfluids at zero temperature. Thus, there is no difference between the superfluid velocity $v_s$ and the center-of-mass velocity $v$ and between the superfluid  density $\rho_s$ and the total mass density $\rho$.  The effective density $\rho_M$, which determines the total transverse force, does not differ from $\rho=\rho_s$. The  Lagrangian in our case  is
\begin{equation}
{\cal L}= \dot{x} P_x -H=  \dot{x} \rho \kappa y- E_v(y)-V(x,y). 
         \label{L2}
\end{equation}
 Ignoring for a while  the defect potential, the original state with a microscopic vortex nucleus  corresponds to $y \approx r_c \approx 0$ and an energy close to zero. After nucleation  there is a vortex
 with zero energy and the coordinate $y$ equal to (see Fig.~\ref{fig1})
\be
y_f={\kappa \over 4\pi v} \ln {\kappa \over v r_c}.
     \ee{yf}
In the classically inaccessible area (underbarrier region) $0<y<y_f$ the energy is positive. Without defects there is no trajectory which crosses the classically
 inaccessible underbarrier region. This is a consequence of translational invariance along the axis $x$. The momentum $ P_x$  is  a constant, and the coordinate $y$ does not vary along any trajectory, even on the complex plane. There is no quantum process which would be able to change its value. So the presence of the defect, which breaks translational invariance, is crucial.  The shape  of the defect is not so essential. It can affect only a pre-exponential factor, which is difficult to calculate anyway.   One can choose the simplest  singular $\delta$-function potential $V(x,y)=- g\delta (x^2+y^2)$.

The nucleation process must start from creation of the vortex nucleus  near the defect at the edge of the film  located a $x=y=0$ in Fig.~\ref{fig1}(a). The energy of the nucleus is on the order of the vortex core energy, which is small compared to logarithmically large vortex energies at finite distance from the film edge. Thus, one can ignore the stage of initial vortex nucleation and consider only the quantum tunneling of the vortex nucleated near the defect through the classically forbidden area. Figure \ref{fig1}(b) shows the trajectory of quantum tunneling in the plane with axes $X=ix$ (imaginary coordinate of the vortex) and $y$  (proportional to  the vortex momentum $P_x$). The trajectory starts at the defect and goes along the line $X=y$ at which the vortex remains at the defect and  interacts with it.  When the trajectory  reaches the point  $y=X=y_f$ it crosses the trajectory with constant $y=y_f$ along which there is no interaction with the defect. The path continues along this new segment of the trajectory until the point $y=y_f$, $x=X=0 $.  Here the path returns from the complex $x$  plane to the real $x$ axis.  The tunneling exponent for such a trajectory is
\begin{eqnarray}
\Gamma  = \frac{2 \mbox{Im} {\cal S}}{\hbar} = -{2 \over \hbar}
\mbox{Im} \left\{\int \limits_{0}^{y_f} P_x(X)\,dX +\int \limits_{y_f}^{0} P_x(y_f)\,dX\right\}
\nonumber \\
= 2\pi n y_f^2=  \frac{\kappa^2n }{8\pi v_s^2} \left(
 \ln \frac{\kappa}{ v_s r_c} \right)^2,
           \label{tun2}
\end{eqnarray}
where $n=\rho/m$ is the particle density.
The probability logarithm $\Gamma$ is roughly equal to the number of particles in the area $y_f^2$ occupied by the velocity field induced by the vortex after nucleation.

\section{Quantum nucleation of a  vortex with mass}

It is interesting to discuss how vortex mass
can affect quantum vortex  nucleation via Magnus tunneling. Muirihead {\em et al.} \cite{Muri} addressed this issue for a two-dimensional vortex near a film edge  on the basis of some estimations
using  a simplified version of the real vortex energy. Later the  semiclassical theory for this case, taking into account the real energy given by \eq{Ham}  was suggested \cite{EBS}. The theory allows to consider the crossover from a massive vortex to a massless vortex nucleated via Magnus tunneling.

For a vortex with mass, the Lagrangian (\ref{L2})  must be modified to
\begin{equation}
{\cal L} = {\mu_v \dot x^2\over 2}+{\mu_v \dot y^2\over 2}+ \kappa\rho_M  \dot x y  - E_v(y),
         \label{Lm2}
\end{equation}
where $\mu_v$ is the mass of the whole vortex (not its unit length) and the energy $E_v(y)$ is given by \eq{Ham} as before.  We have two degrees of freedom, in contrast to one degree of freedom for a massless vortex. Correspondingly, we have two momenta canonically conjugate to coordinates $x$ and $y$:
\begin{equation}
P_x = {\partial {\cal L}\over \partial \dot x}= \mu_v \dot x+\kappa\rho_M y
,\qquad P_y = {\partial {\cal L}\over \partial \dot y} = \mu_v \dot y.
    \end{equation}
The Hamiltonian for the Lagrangian (\ref{Lm2}) is
\bem
{\cal H}={\partial {\cal L}\over \partial \dot x}\dot x+{\partial {\cal L}\over \partial \dot y}\dot y-{\cal L}={\mu_v \dot x^2\over 2 }+{\mu_v \dot y^2\over 2}+E_v(y)
\nonumber \\
={ (P_x-\kappa\rho_M y)^2\over 2  \mu_v}+{P_y^2 \over 2\mu_v}+E_v(y).
   \eem{enMV}
The classical  equations of motion in vortex Cartesian coordinates are
\begin{equation}
\mu_v \ddot x   =-\kappa\rho_M \dot y ,\qquad \mu_v \ddot y   =\kappa\rho_M \dot x-{\partial E_v(y)\over \partial y}.
  \end{equation}
The equations have two integrals. The first integral is the momentum $P_x$, which is conserved because of translational invariance along the axis $x$.  
The second integral is the energy. Now it is possible to find a trajectory for quantum tunneling without defects breaking translational invariance, and one need not retain the energy of interaction with a defect $V(x,y)$. The constant $P_x$ does not mean that the coordinate $y$ cannot vary since $P_x$ depends not only on $y$, but also on $\dot x$.

The relevant  trajectory starts at $y=r_c\approx 0$ and nearly zero energy. \Eq{enMV} shows that the zero energy condition yields the relation connecting $P_x$ with the initial value of $\dot y(0)$ at $y=0$ where   $E_v(y)$ also vanishes:
\begin{equation}
{ P_x^2\over 2  \mu_v}+{\mu_v \dot y(0)^2\over 2 }=0.
        \end{equation}
One can satisfy this condition only at imaginary $\dot y(0)$, and we shall introduce the imaginary time $t=-i\tau$ looking for an underbarrier trajectory. The first integration of the  equations of motion yields
\begin{equation}
 {d y\over d\tau}  ={1\over \mu_v}\sqrt{ 2\mu_v
 E_v(y)+(P_x-\kappa\rho_M y)^2},\qquad
{d x \over d \tau}={P_x-\kappa\rho_M y\over i \mu_v}.
  \end{equation}
The trajectory  starting at $y\approx 0$ ends at the point $y=y_f$ where $E_v(y)=0$ and  the classically accessible  area begins. Eventually the tunneling exponent  is:
\begin{eqnarray}
\Gamma  
= {2 \over \hbar}
\mbox{Im}\left\{ \int \limits_{0}^{x(y_f)} P_x\,dx+\int \limits_{0}^{y_f} P_y\,dy  \right\}= {2 \over \hbar}\mbox{Im}\left\{ \int \limits_{0}^{y_f} \left(P_x{dx\over dy}+P_y\right)\,dy  \right\}
\nonumber \\
=  {2 \over \hbar}\int\limits_0^{y_f}\frac{ 2\mu_vE_v(y)+\kappa\rho_M y(\kappa\rho_M y-P_x)}{\sqrt{ 2\mu_vE_v(y)+(\kappa\rho_M y -P_x)^2}}\,dy.
    \label{GamMV}
\end{eqnarray}
The coordinate $x$ is imaginary along the trajectory, but at the end
of the tunneling trajectory it must become real again. This imposes a condition on the value of the momentum~$P_x$:
\begin{equation}
x(y_f)=\int\limits_0^{y_f}{dx\over dy}dy=i \int\limits_0^{y_f}\frac{ \kappa\rho_M y-P_x}{\sqrt{ 2\mu_vE_v(y)+(\kappa\rho_M y -P_x)^2}}dy=0.
   \label{condPx}
\end{equation}

For a massless vortex ($\mu_v=0$) and $\rho_M=\rho$ \eq{GamMV} yields exactly the same probability exponent as \eq{tun2}, which was obtained using  the complex coordinate but not complex time. In this limit the momentum $P_x$ is cancelled out in \eq{GamMV}. In the opposite limit of very large mass
\begin{equation}
\Gamma={2 \over \hbar }\int\limits_0^{y_f} \sqrt{2\mu_vE_v(y)} \,dy={2 \over \hbar }\int\limits_0^{y_f}\sqrt{2\mu_v\left({\rho \kappa^2 \over  4\pi} \ln{y\over r_c}- v \kappa\rho y\right)}\,dy .
    \end{equation}
This is the standard expression   for tunneling of a particle of mass $\mu_v$ through the potential barrier described by the energy $E_v(y)$.
With logarithmic accuracy, i.e., replacing the logarithm in the integrand by a large constant $\ln{y_f\over r_c}$, one obtains
\begin{equation}
\Gamma   ={4 y_f \over 3\hbar }\sqrt{{\mu_v \rho \kappa^2 \over  2\pi} \ln{y_f\over r_c}}.
    \end{equation}
While at Magnus tunneling the probability logarithm is proportional to $y_f^2$, i.e., the area occupied by the velocity field induced by the vortex, for a massive vortex not subject to the Magnus force it is proportional to  $y_f$.  

Estimating the vortex mass  as the core mass $\mu_v=\pi \rho r_c^2$ the  Magnus tunneling transforms   into  particle-like tunneling at
\be
 {r_c \over y_f}\sqrt{\ln {y_f\over r_c}} >{\rho_M\over \rho}.
      \ee{pM} 
According to \eq{yf},  the ratio $y_f/ r_c$ is on the order of $v_{cr}/v$ (ignoring the logarithmic factor) and large compared to unity for velocities $v$ small compared to the Landau critical velocity $v_{cr}\sim \kappa /r_c$. Moreover, our estimation of the probability exponent is valid as far as the  $y_f/ r_c$ is large. Thus, the left-hand side of the inequality \eq{pM} is small, and the  vortex mass is important for quantum tunneling only at very small $\rho_M /\rho$, i.e., at very strong suppression of the Magnus force in superfluids with broken Galilean invariance.

Above we considered the traditional approach to macroscopic quantum tunneling based on the semiclassical theory for one or two macroscopic degrees of freedom.  It is possible to address the problem   within a more general many-body theoretical framework \cite{Son73,EBS}.  This affected logarithmic factors in expressions for the probability exponent $\Gamma$.

\section{Conclusions}

The Magnus force (one from contributions to the transverse force)  on a vortex appeared  in the pioneer work    on  hydrodynamics of rotating superfluids written by Joe Vinen together with Henry Hall nearly 70 years ago. This force played a crucial role  in Vinen's famous experiment demonstrating quantization of the velocity circulation in  superfluids.  Vinen together with Nozi\`eres \cite{NozVin} connected the 
Magnus force with the Hall effect in superconductors.  He followed  the dispute about the physical nature of the  transverse force  and facilitated its resolution. The transverse force is an important factor in the process of quantum nucleation of vortices, investigations of which were pioneered by Joe Vinen.  

The force is a momentum transferred from one subsystem to another. The most reliable way of its determination is investigation of the momentum balance. There were numerous attempts to determine various components of this force from some general principle referring only to fluid behavior very far from the vortex. The attempt to derive the transverse force from the Berry phase is an example.
The circulation $\kappa_n$ of the normal  velocity,  on which the Berry phase depends, is not a topological charge and
depends on details of interaction of quasiparticles with a vortex at small distances from the vortex.   An information about this interaction is transported to large distances  by the momentum flux.    The Berry phase analysis at large distances itself cannot provide the value of the transverse force without analysis of processes at small distances. The transverse force must be
determined not from the Berry phase, but vice versa: calculation of the transverse force from the momentum balance is necessary for determination of the Berry phase.  

\bmhead{Acknowledgments}

My numerous interactions and discussions with Joe Vinen during the Royal Society Kapitza fellowship in Birmingham and at other various occasions had a great impact on my research work.



\end{document}